\documentclass[10pt,a4paper]{article} 

\usepackage{ifthen}
\newboolean{sv_used}
\setboolean{sv_used}{false}

\usepackage{amsmath}
\usepackage{amsfonts}
\usepackage{amssymb}
\usepackage{bbm} 
\usepackage{mathrsfs} 
\usepackage{graphicx}
\usepackage{color}
\usepackage{epic,rotating} 
\ifthenelse{\boolean{sv_used}}{%
  \usepackage{natbib}
  \bibpunct{(}{)}{;}{a}{,}{,}
  \usepackage{makeidx}
  \usepackage{multicol}        
  \usepackage[bottom]{footmisc}}
{}

 {\begin{quote}\color{red}\begin{sffamily}}%
 {\end{sffamily}\color{black}\end{quote}}
\newcommand{\tmp}[1]{%
  \ifthenelse{\boolean{preview}}{}{\marginpar{\color{red}\begin{sffamily}#1\end{sffamily}\color{black}}}}

\newcommand{\rhs}{right-hand side}

\newcommand{\iid}{i.i.d.}

\newcommand{\ie}{i.e.}

\newcommand{\nset}{\ensuremath{\mathbb{N}}}

\newcommand{\1}{\ensuremath{\mathbbm{1}}}

\newcommand{\eqdef}{\ensuremath{\stackrel{\mathrm{def}}{=}}}

\newcommand{\esssup}[2][]%
{\ifthenelse{\equal{#1}{}}{\left\| #2 \right\|_\infty}{\left\| #2 \right\|_{#1,\infty}}}
\newcommand{\oscnorm}[2][]%
{\ifthenelse{\equal{#1}{}}{\ensuremath{\operatorname{osc}\left(#2\right)}}{\ensuremath{\operatorname{osc}^{#1}\!\left(#2\right)}}}
\newcommand{\essosc}[3][]%
{\ifthenelse{\equal{#1}{}}{\ensuremath{\operatorname{osc}_{#2}{\left(#3\right)}}}{\ensuremath{\operatorname{osc}^{#1}_{#2}\left(#3\right)}}}

\ifthenelse{\boolean{sv_used}}{%
  \spdefaulttheorem{thm}{Theorem}{\bfseries}{\itshape}
  \spdefaulttheorem{prop}{Proposition}{\bfseries}{\itshape}
  \spdefaulttheorem{cor}{Corollary}{\bfseries}{\itshape}
  \spdefaulttheorem{lem}{Lemma}{\bfseries}{\itshape}
  \spdefaulttheorem{defi}{Definition}{\bfseries}{\itshape}
  \spdefaulttheorem{assum}{Assumption}{\bfseries}{\itshape}
  \spdefaulttheorem{rem_duplicate}{Remark}{\bfseries}{\rmfamily}
  \newenvironment{rem}
    {\begin{rem_duplicate}}
    {\smartqed\qed\end{rem_duplicate}}
  \spdefaulttheorem{ex_duplicate}{Example}{\bfseries}{\rmfamily}
  \newenvironment{ex}
    {\begin{ex_duplicate}}
    {\smartqed\qed\end{ex_duplicate}}
  \spdefaulttheorem{algo}{Algorithm}{\bfseries}{\sffamily}
  \spnewtheorem*{proof_duplicate}{Proof}{\itshape}{\rmfamily}
  \renewenvironment{proof}
    {\begin{proof_duplicate}}
    {\smartqed\qed\end{proof_duplicate}}
}{%
  \usepackage{amsthm}
  \newtheorem{thm}{Theorem}
  
  \newtheorem{cor}[thm]{Corollary}
  \newtheorem{lem}[thm]{Lemma}
  
  \newtheorem{assum}[thm]{Assumption}
  \theoremstyle{definition}
  \theoremstyle{definition}
  \theoremstyle{definition}
}
\newenvironment{enum_i}
  {%
  \setlength{\leftmargini}{5ex}%
  \begin{enumerate}}%
  {\end{enumerate}}
  {\end{enumerate}}

\newcommand{\PP}{\ensuremath{\operatorname{P}}}
\newcommand{\PE}{\ensuremath{\operatorname{E}}}
\newcommand{\CPE}[3][]
{\ifthenelse{\equal{#1}{}}{\operatorname{E}\left[\left. #2 \, \right| #3 \right]}{\operatorname{E}_{#1}\left[\left. #2 \, \right | #3 \right]}}
\newcommand{\CPP}[3][]
{\ifthenelse{\equal{#1}{}}{\operatorname{P}\left(\left. #2 \, \right| #3 \right)}{\operatorname{P}_{#1}\left(\left. #2 \, \right | #3 \right)}}
\newcommand{\PVar}{\ensuremath{\operatorname{Var}}}

\newcommand{\gauss}{\ensuremath{\operatorname{N}}}

\newcommand{\unif}{\ensuremath{\operatorname{U}}}

\newcommand{\mult}{\ensuremath{\operatorname{Mult}}}
\renewcommand{\mid}{\,|\,}
\newcommand{\ci}[4][]%
{%
\ifthenelse{\equal{#1}{}}{\ensuremath{#2 \perp\!\!\!\perp #3 \mid #4 }}{\ensuremath{#2 \perp\!\!\!\perp #3 \mid #4 \; \: [#1]}}%
}
\newcommand{\dlim}{\ensuremath{\stackrel{\mathcal{D}}{\longrightarrow}}}
\newcommand{\plim}{\ensuremath{\stackrel{\mathrm{P}}{\longrightarrow}}}

\newcommand{\x}{\ensuremath{x}}
\newcommand{\Xset}{\ensuremath{\mathsf{X}}}

\newcommand{\chunk}[4][]%
{\ifthenelse{\equal{#1}{}}{\ensuremath{{#2}_{#3:#4}}}{\ensuremath{#2^#1}_{#3:#4}}
}

\newcommand{\q}{\ensuremath{q}}
\newcommand{\Xinit}{\ensuremath{\nu}}

\newcommand{\URoot}{\ensuremath{R}}
\newcommand{\UCov}[1][]%
{%
\ifthenelse{\equal{#1}{}}{\URoot \URoot^t}{\URoot_{#1} \URoot^t_{#1}}%
}

\newcommand{\VRoot}{\ensuremath{S}}
\newcommand{\VCov}[1][]%
{%
\ifthenelse{\equal{#1}{}}{\VRoot \VRoot^t}{\VRoot_{#1} \VRoot^t_{#1}}%
}

\newcommand{\LDX}[2]{\ensuremath{L}}

\newcommand{\postdx}[3][]%
{%
\ifthenelse{\equal{#1}{}}{\ensuremath{\psi_{#2|#3}}}{\ensuremath{\psi_{#1,#2|#3}}}%
}

\newcommand{\epostdx}[3][]%
{%
\ifthenelse{\equal{#1}{}}{\ensuremath{\hat{\psi}_{#2|#3}}}{\ensuremath{\hat{\psi}_{#1,#2|#3}}}%
}

\newcommand{\predpx}[3][]%
{%
\ifthenelse{\equal{#1}{}}{\ensuremath{\varphi_{#2|#3}}}{\ensuremath{\varphi_{#1,#2|#3}}}%
}

\newcommand{\filt}[2][]%
{%
\ifthenelse{\equal{#1}{}}{\ensuremath{\phi_{#2}}}{\ensuremath{\phi_{#1,#2}}}%
}
\newcommand{\pred}[3][]%
{%
\ifthenelse{\equal{#1}{}}{\ensuremath{\phi_{#2|#3}}}{\ensuremath{\phi_{#1,#2|#3}}}%
}
\newcommand{\post}[3][]%
{%
\ifthenelse{\equal{#1}{}}{\ensuremath{\phi_{#2|#3}}}{\ensuremath{\phi_{#1,#2|#3}}}%
}
\newcommand{\logl}[2][]%
{%
\ifthenelse{\equal{#1}{}}{\ensuremath{\ell_{#2}}}{\ensuremath{\ell_{#1,#2}}}%
}
\newcommand{\lhood}[2][]%
{%
\ifthenelse{\equal{#1}{}}{\ensuremath{\mathrm{L}_{#2}}}{\ensuremath{\mathrm{L}_{#1,#2}}}%
}
\newcommand{\cc}[2][]%
{%
\ifthenelse{\equal{#1}{}}{\ensuremath{c_{#2}}}{\ensuremath{c_{#1,#2}}}%
}
\newcommand{\forvar}[2][]%
{%
\ifthenelse{\equal{#1}{}}{\ensuremath{\alpha_{#2}}}{\ensuremath{\alpha_{#1,#2}}}%
}
\newcommand{\nforvar}[2][]%
{%
\ifthenelse{\equal{#1}{}}{\ensuremath{\bar{\alpha}_{#2}}}{\ensuremath{\bar{\alpha}_{#1,#2}}}%
}

\newcommand{\BK}[2][]%
{%
\ifthenelse{\equal{#1}{}}{\ensuremath{\mathrm{\mathrm{B}}_{#2}}}{\ensuremath{\mathrm{B}_{#1,#2}}}%
}
\newcommand{\filtfunc}[2][]%
{%
\ifthenelse{\equal{#1}{}}{\ensuremath{\tau_{#2}}}{\ensuremath{\tau_{#1,#2}}}%
}

\ifthenelse{\boolean{sv_used}}{%
  }
{}

\newcommand{\NISE}[4][]%
{%
\ifthenelse{\equal{#1}{}}{\ensuremath{\tilde{#2}^{\scriptstyle \mathrm{IS}}_{#4}\left(#3 \right)}}{\ensuremath{\tilde{#2}^{\scriptstyle \mathrm{IS}}_{#1,#4}\left(#3 \right)}}%
}
\newcommand{\ISE}[4][]%
{%
\ifthenelse{\equal{#1}{}}{\ensuremath{\widehat{#2}^{\scriptstyle  \mathrm{IS}}_{#4} \left(#3 \right)}}{\ensuremath{\widehat{#2}^{\scriptstyle  \mathrm{IS}}_{#1,#4} \left(#3 \right)}}%
}
\newcommand{\SIRE}[4][]%
{%
\ifthenelse{\equal{#1}{}}{\ensuremath{\hat{#2}^{\scriptstyle  \mathrm{SIR}}_{#4} \left(#3 \right)}}{\ensuremath{\hat{#2}^{\scriptstyle  \mathrm{SIR}}_{#1,#4} \left(#3 \right)}}%
}
\newcommand{\MCE}[3]%
{
{\ensuremath{\hat{#1}^{\scriptstyle  \mathrm{MC}}_{#3} \left(#2 \right)}}%
}

\newcommand{\kiss}{\ensuremath{r}}

\newcommand{\XinitIS}{\ensuremath{\rho_0}}

\newcommand{\etpart}[2]{\ensuremath{\tilde{\xi}_{#1}^{#2}}}
\newcommand{\epart}[2]{\ensuremath{\xi_{#1}^{#2}}}
\newcommand{\etwght}[2]{\ensuremath{\tilde{\omega}_{#1}^{#2}}}
\newcommand{\ewght}[2]{\ensuremath{\omega_{#1}^{#2}}}
\newcommand{\ebwght}[2]{\ensuremath{\bar{\omega}_{#1}^{#2}}}

\newcommand{\efilt}[2][]%
{%
\ifthenelse{\equal{#1}{}}{\ensuremath{\hat{\phi}_{#2}}}{\ensuremath{\hat{\phi}_{#1,#2}}}%
}
\newcommand{\epost}[3][]%
{%
\ifthenelse{\equal{#1}{}}{\ensuremath{\hat{\phi}_{#2|#3}}}{\ensuremath{\hat{\phi}_{#1,#2|#3}}}%
}

\renewcommand{\XinitIS}{\ensuremath{{r}_0}}
\renewcommand{\Xinit}{\ensuremath{{\nu}_0}}

\newboolean{arxiv}
\setboolean{arxiv}{true}

\ifthenelse{\boolean{arxiv}}{
\textwidth 160mm
\oddsidemargin -0.5mm
\evensidemargin -0.5mm
\textheight 237mm
\topmargin -15mm}{
\usepackage{times}
\usepackage{ispa/latex8}
\pagestyle{empty}}

\begin{document}

\title{Comparison of Resampling Schemes for Particle Filtering}

\author{Randal Douc\\
Ecole Polytechnique\\
91128 Palaiseau, France\\
{\tt douc} at {\tt cmapx.polytechnique.fr}
\and
Olivier Capp\'{e}\\
Centre National de la Recherche Scientifique\\
46 rue Barrault, 75634 Paris, France\\
{\tt cappe} at {\tt tsi.enst.fr}
\and
Eric Moulines\\
GET T\'{e}l\'{e}com Paris\\
46 rue Barrault, 75634 Paris, France\\
{\tt moulines} at {\tt tsi.enst.fr}}

\maketitle
\thispagestyle{empty}

\begin{abstract}
  This contribution is devoted to the comparison of various resampling
  approaches that have been proposed in the literature on particle filtering.
  It is first shown using simple arguments that the so-called residual and
  stratified methods do yield an improvement over the basic multinomial
  resampling approach. A simple counter-example showing that this property does
  not hold true for systematic resampling is given. Finally, some results on
  the large-sample behavior of the simple bootstrap filter algorithm are given.
  In particular, a central limit theorem is established for the case where
  resampling is performed using the residual approach.
\end{abstract}

\section{Introduction}
The terms \emph{particle filtering} or \emph{Sequential Monte Carlo}
(henceforth abbreviated to SMC), refer to a class of techniques which have
demonstrated a strong potential for signal and image processing applications
\cite{doucet:defreitas:gordon:2001}, \cite{ristic:arulampalam:gordon:2004}.
Schematically, the principle behind sequential Monte Carlo may be viewed
as the combination of two main elements: {\em sequential importance sampling},
which dates back to \cite{mayne:1966,handschin:mayne:1969}, and {\em
  resampling}, whose importance in the context of SMC was first demonstrated by
\cite{gordon:salmond:smith:1993}, based on ideas of \cite{rubin:1987}. In this
contribution, we focus on the second aspect and consider the comparison of
several techniques that have been proposed to implement the resampling step.

To fix the notations, we briefly describe the basic SMC approach known as
\emph{sequential importance sampling with resampling} (or SISR). The algorithm
proceeds as follows:
\begin{itemize}
\item At time 0, draw $m$ particles $\{\epart{0}{i}\}_{1\leq i\leq m}$ from a
  common probability density $\XinitIS$ and compute the associated importance
  weights $\ewght{0}{i} = \Xinit(\epart{0}{i}) g_0(\epart{0}{i}) /
  \XinitIS(\epart{0}{i})$.
\item For successive time indices and for $i=1,\dots,m$, simulate
  $\epart{k+1}{i}$ independently from the past according to a transition
  density function\footnote{In this contribution it is assumed that all
    transition kernels $K(x,dy)$ may be written as $k(x,y) \lambda(dy)$, where
    $\lambda$ is a fixed reference measure (which we usually do not specify);
    $k$ is referred to as a \emph{transition density function}. When $\nu$ is a
    probability density function and $f$ a function, we will use the usual
    notations $\nu(f) = \int \nu(x) f(x) \lambda(dx)$, $kf(x) = \int k(x,x')
    f(x') \lambda(dx')$, $\nu k(x) = \int \nu(x') \lambda(dx') k(x',x)$, and,\\
    \hspace*{4mm} $\nu k f = \int \nu(x) kf(x) \lambda(dx) = \int \nu k(x) f(x)
    \lambda(dx)$\\ \hspace*{32mm} $= \iint \nu(x) k(x,x') f(x') \lambda(dx)
    \lambda(dx')$.}  $\kiss(\epart{k}{i},\cdot)$ and update the weights as
  \[
    \ewght{k+1}{i} = \ewght{k}{i} \q(\epart{k}{i},\epart{k+1}{i})
  g_{k+1}(\epart{k+1}{i}) / \kiss(\epart{k}{i},\epart{k+1}{i})  .
  \]
\end{itemize}
In the context of filtering, $\Xinit$ us the initial distribution of the state
variable, $q$ is the transition density function corresponding to the, possibly
non-linear, state equation (supposed here to be time-homogeneous), and $g_{k}$
is the conditional likelihood of the observation at index $k$ given the
corresponding state, viewed as a function of the state variable. Then, the
self-normalized importance sampling estimator \ifthenelse{\boolean{arxiv}}{
\[
  \sum_{i=1}^m \ewght{k}{i}
f(\epart{k}{i}) \big/ \sum_{j=1}^m \ewght{k}{j}
\]}
{$\sum_{i=1}^m \ewght{k}{i}
f(\epart{k}{i}) \big/ \sum_{j=1}^m \ewght{k}{j}$} is an estimator of the
filtered state moment, that is the expectation of $f$ applied to the
non-observable state variable at time $k$ given all observations up to time
$k$. Not that the choice $\kiss=\q$ is particular in that the weight update
formula then reduces to $\ewght{k+1}{i} = \ewght{k}{i} \,
g_{k+1}(\epart{k+1}{i})$ and thus depends only on the previous weight and new
particle position; when used in conjunction with resampling ideas to be
discussed below this choice ($\kiss=\q$) is known as the \emph{bootstrap
  filter} \cite{gordon:salmond:smith:1993}.

The method sketched so far corresponds to the sequential importance sampling
algorithm, whose drawback is that it becomes unstable as $k$ increase due to
the discrepancy between the weights -- a phenomenon sometimes referred to as
{\em weight degeneracy} \cite[Chapter 7]{cappe:moulines:ryden:2005}.  To
stabilize the algorithm it is necessary to perform resampling sufficiently
often. In the following, we denote by $\{\epart{}{i},\ewght{}{i}\}_{1\leq i\leq
  m}$ the set of particle positions and associated weights at some generic time
index $k$ (which is omitted from our notations) and by $\mathcal{G}^n$ the
$\sigma$-field generated by the generations of particles and weights
up to time $k$, included. We also assume that the weights have already been
normalized, \ie, that $\sum_{i=1}^m \ewght{}{i} = 1$. Resampling consists in
selecting new particle positions and weights
$\{\etpart{}{i},\etwght{}{i}\}_{i=1,\dots,\tilde{M}}$ such that the discrepancy
between the resampled weights $\{\etwght{}{i}\}_{i=1,\dots,\tilde{M}}$ is
reduced. Of course, it is also necessary that the resampled particle system be
as good an approximation to $\{\epart{}{i},\ewght{}{i}\}_{1\leq i\leq m}$ as
possible, in some suitable sense. There are a number of options for performing
resampling and we focus here on the most widely used class of resampling
techniques in which the resampling is random and subject to the constraints
\begin{align}
  & \tilde{M} = n  ,\\
  & \etwght{k}{i} = 1/n  , \\
  & \PE\left[\left. N^i \right| \mathcal{G}^n \right] = n \ewght{k}{i}  , \quad \text{for $i=1,\dots,m$,}
  \label{eq:unbiased}
\end{align}
where $n$ is a non-random integer and $N^i \eqdef \# \{j, 1\leq j \leq n :
\etpart{}{j} = \epart{}{i}\}$ are the particle duplication counts. The third
constraint is sometimes known as the ``unbiasedness'' or ``proper weighting''
condition \cite{liu:chen:1998}. Of course, it is in general most natural to
keep the population size fixed and $n$ is often taken to be equal to $m$. In
some situations however it does make sense to consider resampling scenarios in
which $n$ and $m$ are different, at least for some time indices, and we thus
keep separate notations for these two quantities.

Note that we do not consider here some important resampling algorithms that are
either such that the population size varies (randomly) after resampling
\cite{crisan:delmoral:lyons:1999} or such that the weights are not constrained
to be equal after resampling \cite{fearnhead:clifford:2003}. Our aim with the
present contribution is to complement the results previously published on
resampling in \cite{liu:chen:1998,fearnhead:1998,kuensch:2003,chopin:2004} as
well as to discuss some conjectures.

The rest of the paper is organized as follows: Section~\ref{sec:descr} briefly
describes the four main resampling methods that have been proposed in the
literature which satisfy the constraints mentioned above.
Section~\ref{sec:basic} shows that residual and stratified resampling, as well
as the combination of both, improve over multinomial resampling in the sense of
having lower conditional variance. We also provide a counter-example which
shows that the same property does not hold for systematic resampling, although
its empirical performance is generally found to be close to that of residual
and stratified resampling. Finally, we consider in Section~\ref{sec:clt} the
large sample (\ie, when $n$ increases) behavior of particle filtering methods
which use these various forms of resampling. We are currently able to show
that, in general, central limit theorems hold with the residual resampling
approach, although the target and proposal distributions must satisfy a non
trivial condition.

\section{Description of Resampling Algorithms}
\label{sec:descr}

\subsection{Multinomial Resampling}
The simplest approach to resampling is based on an idea at the core of the
bootstrap method \cite{efron:tibshirani:1993} that consists in drawing,
conditionally upon $\mathcal{G}^n$, the new positions $\{\etpart{}{i}\}_{1\leq
  i\leq n}$ independently from the common point mass distribution $\sum_{j=1}^m
\omega_j \delta_{\xi_j}$. In practice, this is achieved by repeated uses of the
inversion method:
\begin{enumerate}
\item Draw $n$ independent uniforms $\{U^i\}_{1\leq
  i\leq n}$ on the interval $(0,1]$;
\item Set $I^i = D_{\omega}^{\mathrm{inv}}(U^i)$ and $\etpart{}{i} =
  \epart{}{I^i}$, for $i=1,\dots,n$, where $D_{\omega}^{\mathrm{inv}}$ is the
  inverse of the cumulative distribution function associated with the
  (normalized) weights $\{\ewght{}{i}\}_{1\leq i\leq m}$, that is,
  $D_{\omega}^{\mathrm{inv}}(u) = i$ for $u \in ( \sum_{j=1}^{i-1} \ewght{}{j},
  \sum_{j=1}^i \ewght{}{j}]$.  When needed, we will denote by $\epart{}{} :
  \{1,\dots,m\} \to \Xset$ the function such that $\epart{}{}(i) =
  \epart{}{i}$, so that $\etpart{}{i}$ may also be written as $\epart{}{} \circ
  D_{\omega}^{\mathrm{inv}}(U^i)$.
\end{enumerate}
This form of resampling is generally known as \emph{multinomial resampling}
since the duplication counts $N^1, \dots, N^m$ are by definition distributed
according to the multinomial distribution $\mult(n;
\ewght{}{1},\dots,\ewght{}{m})$.

\subsection{Residual Resampling}
Residual resampling, or \emph{remainder resampling}, is mentioned by
\cite{whitley:1994}, \cite{liu:chen:1998} as an efficient means to decrease the
variance due to resampling. In this approach, for $i=1, \dots, m$, we have
\begin{equation}\label{eq:number-particles-residualsampling}
N^i = \left \lfloor n \ewght{}{i}  \right \rfloor + \bar{N}^i ,
\end{equation}
where $\lfloor \, \rfloor$ denotes the integer part and $\bar{N}^1, \dots,
\bar{N}^n$ are distributed according to the multinomial distribution $\mult( n
- R; \ebwght{}{1}, \dots, \ebwght{}{n})$ with $R=\sum_{i=1}^m \lfloor n
\ewght{}{i} \rfloor$ and
\begin{equation}
\label{eq:virtual-multinomialweights}
\ebwght{}{i} = \frac{n \ewght{}{i}  - \lfloor n \ewght{}{i}  \rfloor}{n - R}, \quad i=1, \dots, m.
\end{equation}
This scheme obviously satisfy \eqref{eq:unbiased}. In practice, the multinomial
counts $\bar{N}^1, \dots, \bar{N}^n$ from the residual multinomial distribution
are generated as in the multinomial resampling approach described above.

\subsection{Stratified Resampling}
Stratified resampling -- see \cite{kitagawa:1996} and \cite[Section
5.3]{fearnhead:1998} -- is based on ideas used in survey sampling and consists
in pre-partitioning the $(0,1]$ interval into $n$ disjoint sets, $(0,1] =
(0,1/n] \cup \dots \cup (\{n-1\}/n,1]$. The $U^i$s are then drawn independently
in each of these sub-intervals: $U^i \sim \unif\left(
  \left(\left\{i-1\right\}/n,i/n\right]\right)$, where $\unif([a,b])$ denotes
the uniform distribution on the interval $[a,b]$. Then the inversion method is
used as in multinomial resampling. It is easily checked that, as was the case
for residual sampling, the difference between the duplication count $N^i$ and
its target value $n\ewght{}{i}$ is less than one in absolute value (for all
$i$s). In addition,
\begin{align*}
& \CPE{\sum_{i=1}^n f(\etpart{}{i})}{\mathcal{G}^n} = \CPE{\sum_{i=1}^n f \circ \epart{}{} \circ D^{\mathrm{inv}}_\omega(U^i)}{\mathcal{G}^n} \\
& \quad = n \sum_{i=1}^n \int_{(i-1)/n}^{i/n} f \circ \epart{}{} \circ D^{\mathrm{inv}}_\omega (u)\, du = n \sum_{i=1}^m \ewght{}{i} f(\epart{}{i}) ,
\end{align*}
for all integrable functions $f$, showing that this algorithm also satisfies
\eqref{eq:unbiased}.

\subsection{Systematic Resampling}
Systematic resampling takes the previous method one step further by
deterministically linking all the variables drawn in the sub-intervals. This is
achieved by setting
\[
  U^i = (i-1)/n + U  ,
\]
where $U$ is a \emph{single} random draw from the
$\unif\left(\left(0,1/n\right]\right)$ distribution. Since the $U^i$s generated
this way obviously have the same marginal distribution as those used in the
stratified resampling approach, the method still satisfies \eqref{eq:unbiased}.
It was introduced in the particle filter literature by
\cite{carpenter:clifford:fearnhead:1999} as ``stratified'' sampling but it is
also mentioned by~\cite{whitley:1994} under the name of {\em universal}
sampling. It is often preferred due to its computational simplicity and good
empirical performance. As pointed out by \cite{kuensch:2003} however, it is the
only resampling method for which the resulting particle positions
$\etpart{}{i}$ are no more independent given $\mathcal{G}^n$. Thus, studying its
performance is much harder than for other methods.

A final remark of some importance is that both stratified and systematic
resampling are sensitive to the order in which the particles are ordered: a
simple permutation of the indices of the particles before resampling changes
the distribution of the new resampled set of particles. In contrast, residual
resampling behaves more like the basic multinomial resampling approach in that
it disregards the order in which the particles are numbered.

\section{Basic Properties of Sampling Schemes}
\label{sec:basic}

\subsection{Multinomial Resampling}
For multinomial resampling, the selection indices $I^1, \dots, I^n$ are
conditionally \iid\ given $\mathcal{G}^n$ and thus the conditional
variance is given by
\begin{multline}
  \PVar\left[\left. \frac{1}{n}\sum_{i=1}^n f(\etpart{}{i}) \right| \mathcal{G}^n \right] \\
  = \frac{1}{n}  \left\{ \sum_{i=1}^m \ewght{}{i} f^2(\epart{}{i}) - \left[ \sum_{i=1}^m \ewght{}{i} f(\epart{}{i}) \right]^2 \right\} .
\label{eq:var:multinomial-sampling}
\end{multline}

\subsection{Residual Resampling}
\label{sec:resid:var}
The residual sampling estimator may be decomposed into
\begin{equation}
\label{eq:resampling-estimator}
\frac{1}{n} \sum_{i=1}^n f(\etpart{}{i})= \sum_{i=1}^m \frac{\lfloor n \ewght{}{i}\rfloor}{n} f(\epart{}{i}) + \frac{1}{n} \sum_{i=1}^{n-R} f(\epart{}{\bar{I}^i}) ,
\end{equation}
where $\bar{I}^1, \dots, \bar{I}^{n-R}$ are conditionally independent given
$\mathcal{G}^n$ with distribution $\PP(\bar{I}^i=j\mid\mathcal{G}^n)= \ebwght{}{j}$
for $i=1,\dots,n-R$ and $j=1,\dots,m$. Because the residual resampling
estimator is the sum of one term that, given $\mathcal{G}^n$, is deterministic
and one term that involves conditionally \iid\ draws, the conditional variance
of residual resampling is given by
\begin{align}
\label{eq:variance-residual-sampling}
& \frac{1}{n^2} \PVar \left[\left. \sum_{i=1}^{n-R} f(\epart{}{\bar{I}^i})\right| \mathcal{G}^n \right]
= \frac{n-R}{n^2} \PVar\left[\left.f(\epart{}{\bar{I}^1})\right| \mathcal{G}^n\right]  \\
& \qquad = \frac{1}{n} \sum_{i=1}^m \ewght{}{i} f^2(\epart{}{i}) \nonumber \\
& \qquad \quad - \sum_{i=1}^m \frac{\lfloor n \ewght{}{i} \rfloor}{n^2} f^2(\epart{}{i}) -
\frac{n - R}{n^2} \left\{ \sum_{i=1}^m \ebwght{}{i} f(\epart{}{i}) \right\}^2
. \nonumber
\end{align}
To compare \eqref{eq:variance-residual-sampling}
with~\eqref{eq:var:multinomial-sampling}, first write
$$
\sum_{i=1}^m \ewght{}{i} f(\epart{}{i}) =
\sum_{i=1}^m \frac{\lfloor n \ewght{}{i} \rfloor}{n} f(\epart{}{i})
+ \frac{n - R}{n} \sum_{i=1}^m \ebwght{}{i} f(\epart{}{i})  .
$$
Then note that the sum of the $m$ numbers $\lfloor n \ewght{}{i} \rfloor/n$
plus $(n-R)/n$ equals one, whence this sequence of $m+1$ numbers
can be viewed as a probability distribution. Thus Jensen's inequality
applied to the square of the \rhs\ of the previous display yields
\begin{multline*}
\left\{ \sum_{i=1}^m \ewght{}{i} f(\epart{}{i}) \right\}^2 \\
 \leq \sum_{i=1}^m \frac{\lfloor n \ewght{}{i} \rfloor}{n} f^2(\epart{}{i}) +
\frac{n - R}{n} \left\{ \sum_{i=1}^m \ebwght{}{i} f(\epart{}{i}) \right\}^2
.
\end{multline*}
Combining with \eqref{eq:variance-residual-sampling}, this shows that the
conditional variance of residual sampling is always smaller than that of
multinomial sampling given by \eqref{eq:var:multinomial-sampling}.

\subsection{Stratified Resampling}
Because $U^1, \dots, U^n$ are still conditionally independent given
$\mathcal{G}^n$ for this method,
\begin{align*}
& \PVar\left[\left. \frac{1}{n} \sum_{i=1}^n f(\epart{}{I^i})\right|\mathcal{G}^n\right] = \\
& \frac{1}{n^2} \sum_{i=1}^n \PVar\left[\left. f \circ \epart{}{} \circ D^{\mathrm{inv}}_\omega(U^i)\right|\mathcal{G}^n\right] = \\
& \frac{1}{n} \sum_{i=1}^m \ewght{}{i} f^2(\epart{}{i}) - \frac{1}{n} \sum_{i=1}^n \left[ n \int_{(i-1)/n}^{i/n} f \circ \epart{}{} \circ D^{\mathrm{inv}}_\omega(u) du \right]^2  .
\end{align*}
By Jensen's inequality,
\begin{align*}
& \frac{1}{n} \sum_{i=1}^n \left[ n \int_{(i-1)/n}^{i/n} f \circ \epart{}{} \circ D^{\mathrm{inv}}_\omega(u) du \right]^2 \geq \\ 
& \quad \left[ \sum_{i=1}^n \int_{(i-1)/n}^{i/n} f \circ \epart{}{} \circ D^{\mathrm{inv}}_\omega(u) du \right]^2
 = \left[ \sum_{i=1}^m \ewght{}{i} f(\epart{}{i}) \right]^2 ,
\end{align*}
showing that the conditional variance of stratified sampling is always smaller
than that of multinomial sampling. Note that stratified sampling may be coupled
with the residual sampling method discussed previously: the proof above shows
that using stratified sampling on the $R$ residual indices that are indeed
drawn randomly can then only decrease the conditional variance. It is also
clear that the fact that the conditional variance is reduced does not depend on
the particular choice of the sub-intervals (as being the intervals
$(\{i-1\}/n,i/n]$), more general partitions could be considered as well.

\subsection{Systematic Resampling}
For this last sampling scheme, it is much more complicated to provide a usable
expression of the conditional variance due to all the resampled particles being
(conditionally) dependent \cite{kuensch:2003}. We can however provide a simple
counter-example to the frequently encountered conjecture that systematic
resampling dominates multinomial resampling in terms of conditional variance.

  Consider the case where the initial population of particles
  $\{\epart{}{i}\}_{1\leq i \leq n}$ is composed of the interleaved repetition
  of only two distinct values $x_0$ and $x_1$, with identical multiplicities
  (assuming $n$ to be even). In other words,
\[
  \{\epart{}{i}\}_{1\leq i \leq n} = \{x_0,x_1,x_0,x_1, \dots, x_0,x_1\}  .
\]
We denote by $2\omega/n$ the common value of the normalized weight
$\ewght{}{i}$ associated to the $n/2$ particles $\epart{}{i}$ that satisfy
$\epart{}{i} = x_1$, so that the remaining ones (which are such that
$\epart{}{i} = x_0$) share a common weight of $2(1-\omega)/n$. Without loss of
generality, we assume that $1/2 \leq \omega < 1$ and denote by $|f| = |f(x_1) -
f(x_0)|$.

Under multinomial resampling, \eqref{eq:var:multinomial-sampling} shows that
the conditional variance of the estimate $n^{-1}\sum_{i=1}^n f(\epart{}{i})$
is given by
\begin{equation}
\label{eq::ex_resamp:var_multinom}
\PVar\left[\left. \frac{1}{n}\sum_{i=1}^n f(\etpart{\mathrm{mult}}{i}) \right| \mathcal{G}^n\right] = \frac{1}{n} (1-\omega)\omega |f|^2  .
\end{equation}
In this particular example, it is straightforward to verify that residual and
stratified resampling are equivalent -- which is not the case in general -- and
amount to deterministically setting $n/2$ particles to the value $x_1$ (because
the value $2\omega/n$ is assumed to be larger than $1/n$), whereas the $n/2$
remaining ones are drawn by $n/2$ \emph{conditionally independent} Bernoulli
trials with probability of picking $x_1$ equal to $2\omega-1$. Hence the
conditional variance, for both the residual and stratified schemes, is equal to
$n^{-1}(2\omega -1) (1-\omega) |f|^2$. It is hence always smaller
than~\eqref{eq::ex_resamp:var_multinom}, as expected from the general study of
these two methods. Note that for specific configurations of the weights, such
as when $\omega$ gets close to 0.5, the resampling becomes quasi-deterministic
when using residual or stratified resampling and the improvement over the basic
multinomial scheme becomes all the more significant.

In contrast, systematic resampling also deterministically sets $n/2$ of the
$\etpart{}{i}$ to be equal to $x_1$ but depending on the draw of the initial
shift, {\em all} the $n/2$ remaining particles are either set to $x_1$, with
probability $2\omega-1$, or to $x_0$, with probability $2(1-\omega)$.  Hence
the variance is that of a {\em single} Bernoulli draw scaled by $n/2$, that is,
\[
\PVar\left[\left. \frac{1}{n}\sum_{i=1}^n f(\etpart{\mathrm{syst}}{i}) \right| \mathcal{G}^n\right] =
(\omega - 1/2) (1-\omega) |f|^2  .
\]
note that in this case, the conditional variance of systematic resampling is
not only larger than~\eqref{eq::ex_resamp:var_multinom} for most values of
$\omega$ (except when $\omega$ is very close to $1/2$), but it does not even
decrease to zero as $n$ grows! Clearly, this observation is dependent on the
order in which the initial population of particles is presented. It is easy to
verify (using simulations) that, in this example, systematic resampling becomes
very similar to residual/stratified resampling if the particles are randomly
permuted before resampling. Hence, the above counter-example probably
correspond to a ``rare'' situation. It does however show that systematic
resampling is a variance reduction method which is not as robust as systematic
and residual resampling and also suggest that theoretical study of the behavior
of systematic resampling probably is a very hard task.

\section{Large-Sample Behavior of Resampling}
\label{sec:clt}
We now come to the question of assessing the large sample behavior of particle
filtering methods based on various forms of resampling. The behavior of basic
particle filtering methods when using the multinomial resampling has been
extensively studied in \cite{delmoral:2004}. For reasons of space and
simplicity we only consider here the case of the bootstrap filter (\ie, when
the transition kernel $\q$ of the hidden chain is used as proposal) where
resampling is performed at each time index. In this basic case, each iteration
of the particle filtering algorithm may be decomposed into two successive
steps:
\begin{description}
\item[Prediction] Given the population of unweighted particles at time index
  $k$, $\{\etpart{k}{i}\}_{1\leq i\leq m}$, extend each trajectory
  conditionally independently according to $\epart{k+1}{i} \sim
  \q(\etwght{k}{i},.)$;
\item[Filtering] After computing the weights as
  \[
    \ewght{k+1}{i} = g_{k+1}(\epart{k+1}{i}) / \sum_{j=1}^m g_{k+1}(\epart{k+1}{j})  ,
  \]
  perform resampling to obtain the new unweighted population of particles
  $\{\etpart{k+1}{i}\}_{1\leq i\leq n}$.
\end{description}
The choice of a particular resampling approach does obviously impact only on
the second of these two steps.

To establish central limit theorems for the algorithm above, one can use
repeatedly the two theorems below which are adapted from \cite[Chapter
9]{cappe:moulines:ryden:2005} where the corresponding results are stated under
slightly more general assumptions. The current population of particle is
assumed to satisfy the following assumptions.

\newpage
\begin{assum} \
  \label{assum:recur}
  \begin{enum_i}
  \item \label{item:consist} $\{\epart{}{i}\}_{1\leq i\leq m}$ are consistent
    (in probability) and satisfy a central limit theorem (as $m\to\infty$) for
    a density $\nu$ and all bounded functions $f$, where $\sigma^2(f)$ denotes
    the asymptotic variance, that is,
  \[
  \frac 1m \sum_{i=1}^m f(\epart{}{i}) \plim \nu(f)
  \]
  and
  \[
    \sqrt{m}\left[\frac 1m \sum_{i=1}^m f(\epart{}{i}) -\nu(f)\right] \dlim \gauss(0,\sigma^2(f))
  \]
  for all bounded functions $f$.
  \item The weights are given by
  $\ewght{}{i} = g(\epart{}{i}) / \sum_{j=1}^m g(\epart{}{j})$, where $g(x) =
  \mu(x)/\nu(x)$ for a probability density function $\mu$; $g$ is bounded
  from above and may be known up to a constant only.
  \end{enum_i}
\end{assum}

\begin{thm}
  \label{thm:mutation}
  Under Assumption~\ref{assum:recur}--\ref{item:consist}, new particles $\{\epart{+}{i}\}_{1\leq
    i\leq m}$ distributed conditionally independently under $\epart{+}{i} \sim
  q (\epart{}{i},\cdot)$ are consistent for $\nu \q$ and all bounded functions
  $f$ with asymptotic variance
  \begin{equation}
    \label{eq:mutation:variance}
    \sigma^2_{+}(f) = \nu \left[q f^2 -(qf)^2\right] + \sigma^2(q f)
  \end{equation}
\end{thm}

\begin{thm}
  \label{thm:resampling}
  Under Assumption~\ref{assum:recur}, if \emph{(a)} the resampled particles are
  conditionally independent given $\mathcal{G}^n$, \emph{(b)} $n\to\infty$ with
  $n/m\to\alpha$, and, \emph{(c)}
  \begin{equation}
   \label{eq:lim_cond_var}
   n \PVar\left[\left. \frac{1}{n}\sum_{i=1}^n f(\etpart{}{i})
    \right| \mathcal{G}^n\right] \plim \kappa(f)
  \end{equation}
  that is deterministic, then $\{\etpart{}{i}\}_{1\leq i\leq n}$ are consistent
  and satisfy a central limit theorem for $\mu$ and all bounded functions $f$
  with asymptotic variance
  \begin{equation}
   \label{eq:resampling:variance}
    \tilde{\sigma}^2(f) = \kappa(f) + \alpha \, \sigma^2\left(\frac{\mu}{\nu}[f-\mu(f)]\right)
  \end{equation}
\end{thm}

Following the argument of \cite{kuensch:2003,chopin:2004}, by repeatedly
applying Theorems~\ref{thm:mutation} and~\ref{thm:resampling} one may prove
that the particle filter, when considered at any finite time index $k$, does
satisfy a central limit theorem. The variance formula
in~\eqref{eq:mutation:variance} is a simple instance of the Rao-Blackwell
theorem whereas~\eqref{eq:resampling:variance} shows that the limit of the
conditional variance of resampling gets added to the variance of
(self-normalized or Bayesian) importance sampling scaled by the factor
$\alpha$. This latter factor is interesting as it shows that using $n \ll
m$ may render the variance of the particle estimator almost independent of what
happened in previous steps. This phenomenon should not be over-interpreted
however as it only occurs because the sum is normalized by $n$, and not $m$ (or
$m+n$) which is more connected with the actual number of operations required to
implement the method. Note that the requirement that $g$ be bounded, which is
not very restrictive in the filtering context, may be relaxed --
see~\cite[Chapter 9]{cappe:moulines:ryden:2005} for details.

With multinomial resampling, (\ref{eq:var:multinomial-sampling}) and the
consistency directly implies that $\kappa(f) = \mu(f^2) - [\mu(f)]^2$ that is
the variance under the target density $\mu$. For other resampling schemes
however, showing that~\eqref{eq:lim_cond_var} holds is all but trivial. We
consider in the sequel the case of residual resampling. By
(\ref{eq:variance-residual-sampling}),
\begin{align}
  & n \PVar\left[\left. \frac{1}{n}\sum_{i=1}^n f(\etpart{}{i})
    \right| \mathcal{G}^n\right] \label{eq:ResidVarLim} \\
 & = \sum_{i=1}^m \left(\ewght{}{i} -\frac{\lfloor n \ewght{}{i} \rfloor}{n}\right) f^2(\epart{}{i}) -
\frac{n - R}{n} \left\{ \sum_{i=1}^m \ebwght{}{i} f(\epart{}{i}) \right\}^2 \nonumber \\
 & = \sum_{i=1}^m \left(\ewght{}{i} -\frac{\lfloor n \ewght{}{i} \rfloor}{n}\right) f^2(\epart{}{i}) \nonumber \\
 & \quad - \left. \left\{ \sum_{i=1}^m \left(\ewght{}{i} -\frac{\lfloor n \ewght{}{i} \rfloor}{n}\right) f(\epart{}{i}) \right\}^2\right/ \left(1 -  \sum_{i=1}^m \frac{\lfloor n \ewght{}{i} \rfloor}{n} \right) . \nonumber
\end{align}
Under Assumption~\ref{assum:recur}, for all bounded function $f$, 
$$
 \sum_{i=1}^m \ewght{}{i} f^2(\epart{}{i})=\frac{m^{-1} \sum_{i=1}^m \frac \mu \nu(\epart{}{i})f^2(\epart{}{i})}{m^{-1} \sum_{i=1}^m \frac \mu \nu(\epart{}{i})}\plim \mu(f^2)
$$
and $\sum_{i=1}^m \ewght{}{i} f(\epart{}{i})\plim \mu(f)$. However the case of
sums that involve integer parts cannot be handled similarly and require the
following technical lemma.

\begin{lem}
\label{lem:ResidTechnical}
  Under Assumption~\ref{assum:recur}, if  $n \to \infty$ with $n/m \to \alpha$ and $\mu\left(\1_{\left\{x: \, \alpha\frac \mu \nu(x) \in \nset \right\}}\right)=0$, then  for all bounded function $f$, 
$$
\sum_{i=1}^m \frac{\lfloor n \ewght{}{i} \rfloor}{n} f(\epart{}{i}) \plim  \nu\left\{\frac{1}{\alpha}\left \lfloor\frac{\alpha \mu}{\nu} \right\rfloor f\right \}  .
$$
\end{lem}
\begin{proof}
Recall that  $\ewght{}{i}= g(\epart{}{i}) / \sum_{j=1}^m g(\epart{}{j})$ with $g(x) = \mu(x)/\nu(x)$.
  For any $K \geq 1$, define the set $\mathcal{B}_K= \bigcup_{j=0}^{\infty} [j-1/K,j+1/K]$. 
\begin{align*}
& \sum_{i=1}^{m} \frac{\left \lfloor  n \ewght{}{i} \right \rfloor}{n} \ f(\epart{}{i})\1_{\left\{ \alpha  g(\epart{}{i}) \in (K,\infty)\cup \left((0,K)\cap \mathcal{B}_K \right)\right\}} \\
& \leq \sum_{i=1}^{m} \ewght{}{i} f(\epart{}{i})\1_{\left\{ \alpha  g (\epart{}{i}) \in (K,\infty)\cup \left((0,K)\cap \mathcal{B}_K
\right)\right\}} \\
& \plim \int  f(\x) \1_{\left\{ \alpha g (\x) \in (K,\infty)\cup \left((0,K)\cap \mathcal{B}_K \right)\right\}} \mu(\x) \lambda(d\x)  ,
\end{align*}
where the notation $\1$ stands for the indicator function. The limit on the
\rhs\ of the last display can be made arbitrarily small by taking $K$
sufficiently large because $\int f(\x) \1_{\left\{\alpha g(\x) \in \nset
  \right\}} \mu(d\x) \lambda(d\x)=0$ and $g$ is bounded by
Assumption~\ref{assum:recur}. For any $K \geq 1$, there exists $\eta>0$ such
that
\begin{align*}
  & \1_{\left\{\left| \frac{n}{\sum_{j=1}^m g(\epart{}{i})} - \alpha \right| \leq \eta\right\}} \\
  & \quad \times \1_{\left\{\alpha g(\epart{}{i}) \in (0,K) \setminus
      \mathcal{B}_K \right\}} \left(\left \lfloor n \ewght{}{i} \right \rfloor
    - \left \lfloor \alpha g(\epart{}{i})\right \rfloor\right)= 0 .
\end{align*}
Combining the above with $n/\sum_{j=1}^m g(\epart{}{i}) \plim \alpha$ and
\begin{align*}
& \sum_{i=1}^{m} \frac{\left \lfloor  \alpha g(\epart{}{i}) \right \rfloor}{n} \ f(\epart{}{i}) \1_{\left\{  \alpha  g(\epart{}{i}) \in (0,K) \setminus \mathcal{B}_K
 \right\}}\\
& \quad \plim  \int \frac{\left \lfloor \alpha  g(\x) \right \rfloor}{\alpha}\ f(\x) \1_{\left\{  \alpha  g(\x) \in (0,K) \setminus \mathcal{B}_K \right\}} \nu(\x) \lambda(dx) , 
\end{align*}
yields
\begin{align*}
& \sum_{i=1}^{m} \frac{\left \lfloor n \ewght{}{i}  \right \rfloor}{n} f(\epart{}{i}) \1_{\left\{ \alpha  g(\epart{}{i}) \in (0,K) \setminus \mathcal{B}_K \right\}} \\
& \quad \plim \int \frac{\left \lfloor \alpha  g(\x) \right \rfloor}{\alpha}\ f(\x) \1_{\left\{ \alpha  g(\x) \in (0,K) \setminus \mathcal{B}_K \right\}} \nu(\x)\lambda(d\x). 
\end{align*}
The proof follows by letting $K\to \infty$.
\end{proof}

\begin{cor}
  \label{cor:residual}
  Under Assumption~\ref{assum:recur} and assuming that $\mu\left(\1_{\left\{x:
        \, \alpha\frac \mu \nu (x) \in \nset \right\}}\right)=0$,
\begin{align*}
 & n \PVar\left[\left. \frac{1}{n}\sum_{i=1}^n f(\etpart{}{i}) \right| \mathcal{G}^n\right] \plim \kappa(f) = \\
 & \nu\left\{\left(\frac \mu \nu - \frac{1}{\alpha}\left \lfloor\frac{\alpha \mu}{\nu} \right\rfloor \right)f^2\right\} \\
 & \, - \left. \left[\nu\left\{\left(\frac \mu \nu - \frac{1}{\alpha} \left \lfloor\frac{\alpha \mu}{\nu} \right\rfloor \right)f\right\}\right]^2\right/\left(1 - \nu\left\{ \frac{1}{\alpha}\left \lfloor\frac{\alpha \mu}{\nu} \right\rfloor \right\}\right)
\end{align*}
for the residual sampling method. Hence, the resampled particles satisfy a
central limit theorem with limiting variance given
by~\eqref{eq:resampling:variance}.
\end{cor}
The variance formula given in Corollary~\ref{cor:residual} was first derived in
\cite{chopin:2004} which however lacked a rigorous proof of
Lemma~\ref{lem:ResidTechnical} and the necessity of the support condition --
see \cite{douc:moulines:2005} for a counter-example showing that this condition
is indeed necessary and non-trivially satisfied. Note also that the asymptotic
variance found in Corollary~\ref{cor:residual} is obtained as the (rescaled)
limit of the conditional variance and is thus smaller than in the case where
multinomial resampling is used (see Section~\ref{sec:resid:var}).

\section{Conclusions}
In practical applications of sequential Monte Carlo methods, residual,
stratified, and systematic resampling are generally found to provide comparable
results. Despite the lack of complete theoretical analysis of its behavior,
systematic resampling is often preferred because it is the simplest method to
implement. From a theoretical point of view however only the residual and
stratified resampling methods (as well as the combination of both) may be shown
to dominate the basic multinomial resampling approach, in the sense of having
lower conditional variance for all configurations of the weights. A central
limit theorem as been established for the residual sampling approach. It is
likely that a similar result can be obtained for stratified sampling, based on
Theorem~\ref{thm:resampling}. The situation is however somewhat more involved
in this latter case due to the fact that the new resampled particles, although
still conditionally independent, have a distribution which depend on the order
in which the particles are initially labelled.

\ifthenelse{\boolean{arxiv}}{
\bibliographystyle{plain}
\bibliography{dcm-ispa2005}}{
\bibliographystyle{ispa/latex8}
\bibliography{motherofallbibs}}

\end{document}